\documentclass[sigconf,nonacm]{acmart}
\usepackage[T1]{fontenc}
\usepackage{graphicx}
\usepackage{xifthen}
\usepackage{bussproofs}
\usepackage{listings}
\usepackage{dafny}       
\usepackage{caption}
\usepackage{xcolor}
\usepackage{float}
\usepackage{enumerate}
\usepackage{enumitem}

\newcommand{\met}[1]{\mbox{\bf #1}}
\newcommand{\smet}[1]{\mbox{\small{\bf #1}}}
\newcommand{\pred}[1]{\ensuremath{#1}}
\newcommand{\prepost}[3]{\ensuremath{\{\pred{#1}\} \ \met{#2} \ \{\pred{#3}\}}}

\newcommand{\sorted}[1]{\ensuremath{#1\,\textsl{sorted}}}
\newcommand{\perm}[2]{#1 \sim #2}

\newcommand{\old}{\mbox{\sl old}}
\newcommand{\iif}{\mbox{\sl if}}
\newcommand{\tthen}{\mbox{\sl then}}
\newcommand{\eelse}{\mbox{\sl else}}
\newcommand{\code}[1]{\ensuremath{\mathtt{#1}}}

\lstdefinestyle{myDafny}{
  basicstyle=\scriptsize\ttfamily,
  keywordstyle=\color{blue},
  backgroundcolor=\color{gray!15},
  frame=tb,
  numbers=left,
  stepnumber=1,
  xleftmargin=3.2ex,
  framexleftmargin=3.2ex,
  numbersep=5pt,
  numberstyle=\tiny\color{black},
  commentstyle=\itshape\color{green!40!black},
  emph={assert, by, calc, label},
  emphstyle={\color{red}}
}
\DeclareCaptionFormat{listing} {    
   \parbox{\textwidth}{\hspace{0.1cm}#1#2#3}
}
\title{Derivation and Verification of Array Sorting by Merging, and its Certification in Dafny}

\author{Juan Pablo Carbonell}
\author{José Solsona}
\author{Nora Szasz}
\author{Álvaro Tasistro}
\affiliation{%
  \institution{Universidad ORT Uruguay}
\country{}
}
\email{{carbonell, solsona, szasz, tasistro}@ort.edu.uy}

\begin{document}
\begin{abstract}
We provide full certifications of two versions of merge sort of arrays in the
verification-aware programming language Dafny.
We start by considering schemas for applying the divide-and-conquer or partition method of solution to specifications given by
pre- and post-conditions involving linear arrays. We then derive the merge sort and merging 
algorithms as instances of these schemas, thereby arriving at a fully recursive formulation.
Further, the analysis of the tree of subproblems arising from the partition facilitates the
design of loop invariants that allow to derive a fully iterative version (sometimes called \emph{bottom-up} merge sort) that does not employ a stack.
We show how the use of the provided schemas conveniently conducts the formalization and actual verification in Dafny.
The whole method is also applicable to deriving variants of quicksort, which we sketch.
\end{abstract}              


\maketitle

\section{Introduction.}
This work is directed towards people who  aim at designing and implementing their own programs in a manner such that their purpose and functioning can be explained in plainly well-founded terms.
In this direction, we try to make contributions at two levels: on the one hand, we provide specific merge sort algorithms on arrays, in both recursive and purely iterative versions, along with full certifications, which might be of practical use --despite the algorithms being, at least in principle, extremely well known. Secondly, at a more general, methodological level, we present schemas for algorithm design that apply, in diverse scenarios, the (also well known)  divide-and-conquer technique in a detailed, rigorous manner --an approach that has been proven useful in pursuing the desired kind of certainty at the time of developing and proving the mentioned algorithms.

In particular in this respect, we also contribute an experience with a specification and verification aware programming language, namely Dafny~\cite{leino2010dafny,dafny}, which is a distinguished candidate to become a preferred vehicle for our target audience to express their programs and justifications. In diverse projects, mostly of educational character, we have noticed that Dafny, as well as other tools of its family, demands a very disciplined method of practice so as to properly lead to the desired results. This means that, to our view, the user should approach the tool with a quite elaborate idea of what she wishes to express and why.
The hypothesis has been confirmed in the present experiment, and we share an account of the corresponding process, trying to show in particular how a careful design at an abstract level helps managing the complexity of a subsequent instantiation ---taking on a not at all trivial problem, and aiming at making the whole process replicable. 

Of course, the former does not mean that encoding in Dafny is a redundant task. In this process, we have also been pleased to recognize gaps in our spontaneous, semi-formal reasoning, and helped out of a number of mistakes or omissions. On the other hand, we also shall try to point out spots at which Dafny was at first expected to perform better ---and has instead demanded help in somewhat surprising circumstances.
We also wish all these observations contribute somehow to improving the development of Dafny as a tool, as well as of the general knowledge how to put it into use.

The paper proceeds by introducing a schema of the method of divide-and-conquer on one array (Section~\ref{sec:simple}) that is carefully analyzed and immediately applied to the derivation of recursive merge sort (Section~\ref{sec:mergesortRec}). Then, after discussing some other applications of the schema which include a variant of quicksort (Section~\ref{sec:quicksort}), we go on to develop a recursive merge algorithm (Section~\ref{sec:mergeRec}) based on a schema of divide-and-conquer on two arrays, which we introduce, preceded by another schema on just one array (Sections~\ref{sec:pivot} and~\ref{sec:twoArrays}). This concludes the part on the recursive variant. The rest of the paper is about the iterative version (Section~\ref{sec:mergesortIt}), which derives from an analysis of the ``tree of calls" produced in general by the divide-and-conquer schemas from Section~\ref{sec:rec2it}.
We conclude commenting on related and further work in Section~\ref{sec:conclusions}. 

\section{Divide-and-Conquer: simple schema.}
\label{sec:simple}

We start by presenting a simple schema for developing algorithms by application of one form of the divide-and-conquer method. 
The prime intention is to employ the schema for actually guiding the development of merge sort on arrays, together with its formal verification. We shall also show its applicability to other problems, including a (sketch of) development of quicksort.


We shall therefore be concerned with linear arrays, i.e., methods acting on them or computing values from them.
To begin with, consider problems centered upon \emph{one} array $a$ of length $N$.
The type of the elements of the array is immaterial.
We therefore have a problem (specification) of writing a method \met{\underline{M}}, one of whose parameters is $a$, and subject to precondition \pred{\underline{Q}}  and post-condition \pred{\underline{R}}, which we write:
\[\prepost{\underline{Q}}{\met{\underline{M}}}{\underline{R}}\]

\hfill

\noindent Here, \pred{\underline{Q}}  and  \pred{\underline{R}} are predicates on $a$, besides other variables relevant to the problem. Actually, the present schema will only depend on $\pred{\underline{Q}} \ \Rightarrow N\geq 0$, i.e. so as to impose no restriction on the length of the array.
Here and in the sequel we omit as many arguments to predicates and methods as seen fit for readability. In particular, we shall systematically omit (the global array) $a$.
Now the whole idea of divide-and-conquer is that the solution 
performs the following steps: 

\begin{description}
    \item[Divide.] Conveniently split the array into several sub-arrays (slices).
    \item[Conquer.] Call itself on each of the slices. 
    \item [Combine.] Combine the outcomes so that \pred{\underline{R}} gets satisfied.  \\
\end{description}
\noindent We readily notice that we should then be able to operate on {\em slices} of $a$, which calls for a generalization, consisting in parameterizing the whole problem in the slice in question. 
This leads us to specifications 
\begin{center}
\[\prepost{Q_{l,r}}{M$_{l,r}$}{R_{l,r}}\]    
\end{center}
where:\begin{enumerate}
    \item We use slices $a[m..n)$, i.e. of length $n-m$, where generally we shall have $0 \leq m \leq n \leq N$,
    \item $l$ and $r$ are the left and right limits of the parameter slice, and are \emph{value-} parameters to \met{M}, i.e. not mutable therein,
    \item $\pred{Q_{l,r}} \ \Rightarrow 0 \ \leq l \leq r \leq N$,
    \item $\pred{\underline{Q}} \ \Rightarrow \  \pred{Q}_{0,N}$, and \label{cond:Q}
    \item $\pred{R}_{0,N} \ \Rightarrow \ \pred{\underline{R}}$. \label{cond:R}\\
\end{enumerate}
The original problem can then be solved in terms of a solution for the generalization, by the simple call
\[\underline{\met{M}} \ \triangleq \ \met{M} _{0,N}, \]

\noindent since condition~(\ref{cond:Q}) makes the generalized pre-condition \pred{Q} at the whole array a relaxation of the originally given \pred{\underline{Q}}, as the same time as that the converse holds between the generalized post-condition at the whole array and the original post-condition 
\pred{\underline{R}}, because of~(\ref{cond:R}).

Actually, we should and shall strengthen the post-condition by requiring that \met{M}, when operating on a slice of $a$ does not affect the rest of $a$, so as to make the diverse recursive calls (sub-processes of the solution) independent of each other --- specifically, interference-free.
We therefore write our specifications as:

\[\prepost{Q_{l,r}}{M$_{l,r}$}{R_{l,r} \land S_{l,r}}\]

\noindent where, for formulating \pred{S}, we need to refer to the state of elements of the array \emph{before} the execution of \met{M}, which we do by borrowing from Dafny the notation \old. 

\noindent  Then
\[\pred{S}_{l,r} \Rightarrow  (\forall i: 0 \leq i < l \ || \ r \leq i < N) \ a[i]=\old(a[i]),\]

\noindent so that $\pred{S}_{l,r}$ can be read "not affecting $a$ outside of the slice $[l..r)$".
Note the equivalent (and preferred) definition of $\pred{S}_{l,r}$ as:
 \[\pred{S}_{l,r} \Rightarrow  a[0..l) = \old(a[0..l))  \ \land  \  a[r..N) = \old(a[r..N)).\]

Now, proceeding on, it is obvious that the recursive decomposition must yield subproblems of lesser size, and that some base case must be provided in which decomposition is not feasible or convenient.
So, we introduce a predicate \pred{B} on $l$ and $r$ which characterizes such basic cases. Therefore, $\neg  \pred{B}$ will characterize the cases in which decomposition may proceed. Two observations are pertinent at this point:

\begin{enumerate}
    \item[(a)] We shall here depict a form of {\em binary} divide-and-conquer.
    \item[(b)] We shall partition the slice $[l..r)$ into two by an intermediate index $m$. Then, the recurrence will proceed on these two parts. This stands in contrast to a so-called pivot-based partitioning, in which an intermediate element of the array is selected which does not participate in any of the slices to which the recurrence will apply.
    It is relevant to consider pivot-based partitioning too, and we shall do so later. Among its applications is prominent a well-known variant of quicksort. We shall show another application too.
\end{enumerate}


Having made these observations, a little reflection will show that it must be guaranteed that \emph{both} slices arising from the partition must be shorter than the original one, which enforces them to be non-empty. The latter implies the existence of at least one element in each originating slice, hence the size $r-l$ of the original slice is at least $2$.
Therefore, we must have 
    \begin{equation}\label{eqn:r-l}
    \neg  \pred{B}_{l,r} \ \Rightarrow \ r-l > 1 
    \end{equation}
    
\noindent This completes the analysis of the base case, i.e., we must have $B$ satisfying~(\ref{eqn:r-l}) and solve:
\[\textbf{(Base) } \prepost{Q_{l,r} \ \land \ B_{l,r}}{E$_{l,r}$}{R_{l,r} \wedge S_{l,r}}.\]  

We have already mentioned the partition step, yielding the intermediate index (to be called $m$). We here specify this step in a quite general manner, namely
\[\textbf{(Divide) } \prepost{Q_{l,r} \ \land \neg B_{l,r}}{D$_{l,m,r}$}{P_{l,m,r}},\]

\noindent where: \begin{enumerate}
    \item $P_{l,m,r} \Rightarrow l < m < r$, as already explained. This guarantees a well-founded recursion with variant $r-l$ (the length of the slice), with the recursive calls being of course $\met{M}_{l,m}$ and  $\met{M}_{m,r}$.\label{cond:P1}
    \item $P_{l,m,r} \Rightarrow  Q_{l,m} \land Q_{m,r}$, 
    thereby enforcing that the pre-condition predicate \pred{Q} be established for the two originating slices, so that the same problem in the beginning can be considered on them.\label{cond:P2}
    \item The condition \pred{P} should be preserved by the recursive calls, so as to allow for a provably correct sequential execution of the latter. This is enforced by \begin{enumerate}
        \item $R_{l, m} \land S_{l, m} \Rightarrow (P_{(l,m,r)\uparrow\smet{M}_{l,m}} \Rightarrow P_{l,m,r})$\label{cond:RyS1}
        \item $R_{m, r} \land S_{m, r} \Rightarrow (P_{(l,m,r)\uparrow\smet{M}_{m,r}} \Rightarrow P_{l,m,r})$\label{cond:RyS2},
    \end{enumerate}
\noindent where we use the notation $\overline{x} \uparrow \met{M}$ to denote the state of variables $\overline{x}$ before the (preceding) execution of method \met{M}.
\end{enumerate}



The recursive calls $\met{M}_{l,m}$ and  $\met{M}_{m,r}$ can now be carried out in parallel (notice we have avoided interference) or sequentially, as we shall consider presently. 
Notice that they establish \pred{P_{l,m,r} \land R_{l,m} \land S_{l,m} \land R_{m,r} \land S_{m,r} }, as shown in the following deduction (program tableau):
\begin{align*}
& \{P_{l, m ,r}\}\ (\text{as arising from the execution of the divide phase } \met{D$_{l,m,r}$}) \\ 
& \{\pred{l < m < r \ \land \ Q_{l,m} \ \land \ Q_{m,r}}\} \ (\text{by (\ref{cond:P1})  and~(\ref{cond:P2}) above on } \pred{P})\\
&\{Q_{l,m}\}\\
& \hspace*{2em}\met{M$_{l,m}$}\\
& \{\pred{R_{l,m} \ \land \ S_{l,m}}\} \ \ (\text{by induction hypothesis})\\
& \{\pred{R_{l,m} \ \land \ P_{l,m,r}}\} \ \ (\text{by the preservation condition~(\ref{cond:RyS1}}))\\
& \{P_{l, m ,r}\}\\ 
& \{\pred{l < m < r \ \land \ Q_{l,m} \ \land \ Q_{m,r}}\} \ (\text{again, by (\ref{cond:P1}) and~(\ref{cond:P2}) on } \pred{P})\\
& \{Q_{m, r}\}\\
& \hspace*{2em}\met{M$_{m,r}$}\\
& \{\pred{R_{m,r} \land \ S_{m,r}}\} \ (\text{by induction hypothesis})\\
& \{P_{l,m,r} \land \pred{R_{l,m} \land S_{l,m} \land \ R_{m,r} \land S_{m,r}}\},\\
\end{align*}
\noindent where the last step is justified by
\begin{itemize}
    \item \pred{S_{m,r}} and $l<m$ imply $a[l..m) = (a[l..m)\uparrow\smet{M}_{m,r}))$ 
    ---i.e. nothing is altered in $a$ outside the slice $[m..r)$---, and therefore $R_{l,m} \land S_{l,m}$ is preserved.
    \item Similarly, condition~(\ref{cond:RyS2}) above guarantees the preservation of \pred{P} by the second recursive call.\\
\end{itemize}
This last observation makes the partition condition \pred{P} available as a pre-condition to the combination phase, which is just natural. We therefore have, finally:
\[\textbf{(Combine) } \prepost{P_{l,m,r} \land R_{l,m} \land S_{l,m} \land \ R_{m,r} \land S_{m,r}}{C$_{l,m,r}$}{R_{l,r} \ \land \  S_{l,r}},\]  


\noindent and the program scheme that results is, therefore:  
\begin{equation*}
\begin{array}{ccl}
\met{M}_{l,r}  &  \triangleq  & \iif \  \pred{B_{l,r}} \  \tthen \  \met{E$_{l,r}$}\\
& & \eelse\ \ \met{D$_{l,m,r}$}\\
& & \hspace*{2.2em}\met{M$_{l,m}$} \\
& & \hspace*{2.2em} \met{M$_{m,r}$} \\
& & \hspace*{2.2em} \met{C$_{l,m,r}$}. \\\\
\end{array}
\end{equation*}

\section{Merge Sort, Recursively.}
\label{sec:mergesortRec}

Merge sort is a comparison-based sorting algorithm that follows the divide-and-conquer paradigm. It recursively divides an array into two halves, sorts each half, and then merges the sorted halves to produce a fully ordered array. This method ensures a worst-case time complexity of $\mathcal{O}(n \log n)$, making it efficient even for large datasets~\cite{knuth1998, cormen2022}.

In order to obtain a first version of that algorithm, we proceed now to instantiate the preceding schema by providing the various necessary components.
For simplicity, the type of the elements of the array is just taken to be \code{int} and, besides, the ordinary order relation is used.
The specification of sorting obtains by making:
\[\underline{\pred{Q}} \ \triangleq \ N \geq 0, \]
\[\underline{\pred{R}} \ \triangleq \ \sorted{a} \ \land \ \perm{a}{\old{(a)}}, \]

\noindent where we use \textsl{sorted} with the obvious meaning\footnote{Just for simplicity we will impose ascending order.}
and $\sim$ for the permutation relation. 
Here in the main text we do not make any notational difference between arrays and the sequences of values therein\footnote{But it will have to be made when passing on to Dafny code.}.
Passing on to the generalized method working on slices, we have:
\begin{align*}
\pred{Q_{l,r}} \triangleq&\,\ 0 \leq l \leq r \leq N, \\
\pred{R_{l,r}} \triangleq&\,\ \sorted{a[l..r)} \ \land \ a[l..r) \sim \old(a[l..r)), \\
\pred{S_{l,r}} \triangleq&\,\ a[0..l) = \old(a[0..l)) \ \land \ a[r..N) = \old(a[r..N)), \\
\pred{B_{l,r}} \triangleq &\,\ r-l < 2, \text{ which gives immediately } \ \met{E$_{l,r}$} \triangleq \text{\sl skip}, \\
\pred{P_{l,m,r}} \triangleq &\,\ 0 \leq l < m < r \leq N, \\
\met{D$_{l,m,r}$} \triangleq &\,\ m:= (r+l) / 2, \\
\met{C$_{l,m,r}$} \triangleq &\,\ \met{merge} ,
\end{align*}

\noindent and therefore we derive the specification:
\[\prepost{P_{l,m,r} \land R_{l,m} \land S_{l,m} \land \ R_{m,r} \land S_{m,r}}{merge}{R_{l,r} \ \land \  S_{l,r}}.\]
\noindent It can readily be checked that the provisos dictated in the preceding section for the various components are indeed satisfied. We list them for facilitating the check:

\begin{itemize}
    \item $\pred{Q_{l,r}} \ \Rightarrow 0 \ \leq l \leq r \leq N$,
    \item $\pred{\underline{Q}} \ \Rightarrow \  \pred{Q}_{0,N}$, 
    \item $\pred{R}_{0,N} \ \Rightarrow \ \pred{\underline{R}} $,
    \item $\pred{S}_{l,r} \Rightarrow  a[0..l) = \old(a[0..l))  \ \land  \  a[r..N) = \old(a[r..N)),$
    \item $\neg  \pred{B}_{l,r} \ \Rightarrow \ r-l > 1,$
    \item $P_{l,m,r} \Rightarrow l < m < r,$
    \item $P_{l,m,r} \Rightarrow  Q_{l,m} \land Q_{m,r}$,
    \item\begin{enumerate}[leftmargin=0.5cm]
        \item[(a)] $R_{l, m} \land S_{l, m} \Rightarrow (P_{(l,m,r)\uparrow\smet{M}_{l,m}} \Rightarrow P_{l,m,r})$
        \item[(b)] $R_{m, r} \land S_{m, r} \Rightarrow (P_{(l,m,r)\uparrow\smet{M}_{m,r}} \Rightarrow P_{l,m,r})$,
        \end{enumerate}
    \item $\prepost{Q_{l,r} \ \land \ B_{l,r}}{E$_{l,r}$}{R_{l,r} \wedge S_{l,r}},$
    \item $\prepost{Q_{l,r} \ \land \neg B_{l,r}}{D$_{l,m,r}$}{P_{l,m,r}},$
    \item $\prepost{P_{l,m,r} \land R_{l,m} \land S_{l,m} \land \ R_{m,r} \land S_{m,r}}{C$_{l,m,r}$}{R_{l,r} \ \land \  S_{l,r}}.$\\
\end{itemize}

For instance, the double proviso on the preservation of \pred{P} is satisfied simply because, as already noted, the value parameters $l,m,r$ cannot be altered by the corresponding calls of \met{M}.
On the other hand, the specification of \met{merge} can be simplified to:

\hfill

$\{0 \leq l < m < r \leq N \land \sorted{a[l..m)} \,\land \ \sorted{a[m..r)}\}$

$\met{merge}$

$\{\sorted{a{[l..r)}} \,\land \ a[l..r) \sim \old(a[l..r)) \land \  S_{l,r}\}$,

\hfill

\noindent simply because we are  relaxing the pre-condition.

\hfill

We shall shortly derive \met{merge} from another schema of divide-and-conquer.
Meanwhile, let us just show how the derived merge sort can be encoded and verified in Dafny from just the specification of \met{merge}. 
For understanding the code below, first have in mind that the notation of slices in Dafny uses \texttt{]} (instead of \texttt{)}) for expressing right-open-ended intervals.
The method \texttt{merge\_sort}, as well as the specification and the executable part of the auxiliary generalization \texttt{merge\_sort'} are then direct after the preceding discussion. 
The same happens with the specification of \texttt{merge} at the bottom.
In \texttt{merge\_sort'}, Dafny requires help for proving the post-condition in line 12, i.e. of the fact that the method renders a permutation of the given slice. The proof is given by the two \texttt{calc} statements appearing (slightly indented to the right) in lines 21 to 33 and 36 to 49.


\begin{lstlisting}[basewidth={.55em}]
method merge_sort (a:array<int>)
 modifies a
 ensures sorted (a[0..a.Length])
 ensures perm (a[0..a.Length], old(a[0..a.Length]))
{
 merge_sort' (a, 0, a.Length) ;
}

method merge_sort' (a:array<int>, l:int, r:int)
 modifies a
 requires 0 <= l <= r <= a.Length
 ensures sorted (a[l..r])
 ensures perm (a[l..r], old (a[l..r]))
 ensures a[0..l] == old (a[0..l])
 ensures a[r..a.Length] == old (a[r..a.Length])
 decreases r - l
{
 if r-l < 2 { }
 else {
  var m := (r+l)/2 ;
  merge_sort' (a, l, m) ;
   calc {
    a[m..a.Length] == old(a[m..a.Length]);
    ==> { sub_eq(a[..],old(a[..]),m,r,a.Length); }
    a[m..r] == old(a[m..r]);
    ==> 
    perm (a[m..r],old(a[m..r]));
    ==> { 
      assert a[l..r] == a[l..m] + a[m..r];
      assert old(a[l..r]) == old(a[l..m])+old(a[m..r]);
      perm_sum(a[l..m],old(a[l..m]),a[m..r],old(a[m..r]));
     }
    perm(a[l..r], old(a[l..r]));
   }
   label bfr:
  merge_sort' (a, m, r) ;
   calc {
    a[0..m] == old@bfr(a[0..m]);
    ==> { sub_eq(a[..],old@bfr(a[..]),0,l,m); }
    a[l..m] == old@bfr(a[l..m]);
    ==> 
    perm (a[l..m],old@bfr(a[l..m]));
    ==> { 
      assert 0 <= l <= m < r <= a.Length;
      assert a[l..r] == a[l..m] + a[m..r];
      assert old@bfr(a[l..r]) == old@bfr(a[l..m])+old@bfr(a[m..r]);
      perm_sum(a[l..m],old(a[l..m]),a[m..r],old@bfr(a[m..r]));
     }
    perm(a[l..r], old@bfr(a[l..r]));
   }
  merge (a, l, m, r) ;
 }
}

method {:axiom} merge (a : array<int>, l : int, m : int, r : int)
 modifies a
 requires 0 <= l < m < r <= a.Length
 requires sorted (a[l..m]) && sorted (a[m..r])
 ensures sorted (a[l..r])
 ensures perm (a[l..r], old(a[l..r]))
 ensures a[0..l] == old (a[0..l]) && 
         a[r..a.Length] == old (a[r..a.Length])
\end{lstlisting}

\hfill

\noindent 
We now describe these proofs: starting form the bottom of the method (line 48) in upwards direction, we first prove that the slice in question (\texttt{a[l..r]}) is a permutation of the same before the second recursive call. Then we do a similar thing to go through the first recursive call (line 31 upwards).
Looking for instance at the bottom proof, we read the desired conclusion at line 48. This is arrived at (conceptually) by splitting the interval \texttt{[l..r)} into \texttt{[l..m)} and \texttt{[m..r)} ---i.e. by showing that the slices at those two sub-intervals are permutations of themselves before the recursive call. 
Now for \texttt{[l..m)} this desired condition is a consequence of the non-interference post-condition of the recursive call in question ---i.e. that nothing happens outside the interval \texttt{[m..r)}---, and is explicitly established in lines 37 to 41:
there we start from the non-interference post-condition to 35 and then need to proceed by using a lemma (\texttt{sub\_eq}) for showing that equality of two slices implies equality of given suffix thereof (actually the same lemma is used in the upper proof for prefixes, and has a straightforward, minimal proof).
Then the step at 41 appears necessary, making it explicit use of the fact that permutation is reflexive (it needs no help for checking).
On the other hand, that the permutation condition holds for the slice at \texttt{[m..r)} is just an induction hypothesis, and actually needs not be made explicit.
So the key step is that from 42 to 47. Its justification happens to be necessary for the proof to check: we have to establish basic facts about concatenation of slices by the \texttt{assert} statements at 44 and 45, and then call a lemma (\texttt{perm\_sum}) to the effect that the permutation relation is preserved by sum (concatenation) of slices. The proof of this lemma is automatic.

\section{Further Application of the Schema: Sketch of Quicksort.}
\label{sec:quicksort}

Another application of the schema results in a variant of quicksort.
The specification is of course the same, as are the supplementary predicate \pred{S} and the base case predicate \pred{B}. The distinction shows up in the partition predicate \pred{P}, which must require the left-side slice to be (uniformly) less than or equal to the right-side one. Consequently, the method $\met{D}$ must implement such partition, and it can be derived from the preceding schema in the following way: a value in the slice is picked and then the slice to be partitioned is divided into two halves to each of which division with respect to the chosen value is applied. Then a method of slice swapping is required ---which can in turn be approached under the schema of divide-and-conquer on two arrays (slices) to be presented here later.
On the other hand, the final combination method \met{C} of quicksort becomes trivial (i.e. is just \textsl{skip}).\\


\section{Divide-and-Conquer: Schema with a Pivot.}
\label{sec:pivot}

We have already mentioned the possibility of performing the division phase by selecting an element of the slice that will not participate in any of the slices to which the recurrence applies. 
A first observation concerning this alternative is that it demands the partitioned slice to be just non-empty ---and not of length at least two, as in the former method. 
As we shall see below, this detail actually separates the two approaches, making them applicable (at least, in a direct, not awkward manner) to different circumstances.

First of all, the schema with a pivot is not very good for merge sort. Indeed, the combine phase would have to merge the two slices arising form the partition and the recursive calls \emph{together} with the pivot, thus requiring a more complicated algorithm.

 Besides, on the other hand, our first simple schema is not quite good for approaching the merge of two sorted slices. In this case, a divide-and-conquer approach would partition one of the slices (call it the \emph{principal} one) as above, i.e. into two halves. Then it would practice a subsidiary partition of the other (\emph{secondary}) array, and merge together the left-hand (lesser) parts and, separately, the right-hand (greater) parts. The partition of the secondary array should have to be performed using a discriminating value, kind of watershed between the lesser and greater parts of the partition, which ought in turn to be the half point of the principal array ---for reasons of efficiency. Now, in the case in which the principal array is of length one, our former method of partition is not applicable, and thereby we would be forced to making use of an algorithm of insertion of this solitary value into the secondary array ---a process solely needed in this special case. 

The latter observation motivates the development of a different schema as framework for the problem of merging, i.e. one in which there are both the principal and secondary arrays to begin with, and where a pivot is preferably used for partitioning the principal one. 
Now, before entering the description of such a \emph{two-arrays} schema, we wish to look more closely at the pivot-guided one for just one array ---in order to manage complexity in a gradual manner.
Now, the adjustment to be made to the former schema is actually quite minor, as we see presently. 

We have, to begin with, that the non-base case, i.e. when the partition is to take place, must imply just the existence of the pivot. Therefore, if we are working on slices
$a[l..r)$, then we must have 
\[\neg  \pred{B}_{l,r} \ \Rightarrow \ r-l > 0.\]

\noindent Then we shall select an intermediate point of the (non-empty) interval, which obtains if we demand the partition predicate \pred{P} to satisfy
\[P_{l,m,r} \Rightarrow l \leq m < r,\]
and we are ready to proceed recursively on the parts if, besides:
\[P_{l,m,r} \Rightarrow  Q_{l,m} \land Q_{m+1,r}.\]

\noindent That is to say that the recursive calls will proceed on the slices $a[l..m)$ and $a[m+1..r)$, so that the resulting program scheme is

\begin{equation*}
\begin{array}{ccl}
\met{M}_{l,r}  &  \triangleq  & \iif \  \pred{B_{l,r}} \  \tthen \  \met{E$_{l,r}$}\\
& & \eelse\ \ \met{D$_{l,m,r}$}\\
& & \hspace*{2.2em}\met{M$_{l,m}$} \\
& & \hspace*{2.2em} \met{M$_{m+1,r}$} \\
& & \hspace*{2.2em} \met{C$_{l,m,r}.$} \\
\end{array}
\end{equation*}

Accordingly, the preservation of the partition predicate demands:

\begin{enumerate}
        \item[(a)] $R_{l, m} \land S_{l, m} \Rightarrow (P_{(l,m,r)\uparrow\smet{M}_{l,m}} \Rightarrow P_{l,m,r})$
        \item[(b)] $R_{m+1, r} \land S_{m+1, r} \Rightarrow (P_{(l,m,r)\uparrow\smet{M}_{m+1,r}} \Rightarrow P_{l,m,r})$.
\end{enumerate}

For instance, a usual implementation of quicksort would use the present schema with $P$ imposing the left-hand slice $a[l..m)$ to be (uniformly) lesser than or equal to $a[m]$, and this one to be less than ot equal to each element of $a[m+1..r)$.

\section{Divide-and-Conquer: Schema with two arrays.}
\label{sec:twoArrays}
We wish now to treat multiple structures altogether, specifically along the lines of the \emph{two arrays} situation posed in the preceding section relating the merge problem. 

We therefore consider a principal array slice $a[l..r)$, together with a secondary slice $b[l'..r')$. The idea we wish to capture is that of working recursively on a partition of the principal slice that induces a subsidiary partition of $b$. The partition of $a$ could be carried out along the lines of either of the schemas so far analyzed, but for convenience for the application that we have in mind, namely the merge algorithm, we now choose the divide-and-conquer with pivot.

As a consequence, we shall have the already identified and characterized the \emph{base} condition \pred{B} for the principal slice. Now, a similar condition \pred{B'} for the secondary structure is also convenient, for it makes the design of the executable code more flexible. We can actually move on to giving the program scheme:

\begin{equation*}
\begin{array}{ccl}
\met{M}_{l,r,l',r'}  &  \triangleq  & \iif \  \pred{B_{l,r}} \  \tthen \  \met{E}_{l,r,l',r'}\\
& & \eelse\ \iif\ \pred{B'_{l',r'}} \tthen\ \met{E'}_{l,r,l',r'}\\
& & \eelse\ \ \met{D$_{m,l'_{1},r'_{1},l'_{2},r'_{2}}$}\\
& & \hspace*{2.2em}\met{M$_{l,m,l'_{1},r'_{1}}$} \\
& & \hspace*{2.2em} \met{M$_{m+1,r,l'_{2},r'_{2}}$} \\
& & \hspace*{2.2em} \met{C$_{l,m,r,l'_{1},r'_{1},l'_{2},r'_{2}.}$} \\
\end{array}
\end{equation*}

Some comments are in point now.
First of all, notice that the recursion proceeds on the principal array $a$, and has as variant the size of this slice. This follows from the use of the two first parameters to \met{M} and what we already know on the base condition \pred{B}.
On the other hand, the division method \met{D} produces the intermediate point $m$ of the principal slice, as well as a (binary) partition of the secondary structure $b$ into slices $[l'_{1}..r'_{1})$ and $[l'_{2}..r'_{2})$. Since we are considering binary partition of the principal slice, and generally a binary structure of the divide-and-conquer schema, this characterization seems fit.

As a consequence, we have specifications 
\[\prepost{Q_{l,r,l',r'}}{M$_{l,r,l',r'}$}{R_{l,r,l',r'} \land S_{l,r,l',r'}},\] 
where:
\begin{itemize}
    \item $\pred{Q_{l,r,l',r'}} \ \Rightarrow 0 \leq l \leq r \leq N \land 0  \leq l' \leq r' \leq N'$, for $N'$ the length of $b$,
    \item $\pred{S}_{l,r,l',r'} \Rightarrow  
         \begin{array}[t]{l}
             \hspace{8pt} a[0..l) = \old(a[0..l))  \\
             \land\ a[r..N) = \old(a[r..N)) \\
             \land\ b[0..l') = \old(b[0..l')) \\
             \land\ b[r'..N) = \old(b[r'..N)),
         \end{array}$\\
       i.e. an extended non-interference condition.
\end{itemize}
Besides, we introduce, similarly as for the original schema, base condition $B$ and partition predicate $P$ such that:
\begin{itemize}
    \item $\neg  \pred{B}_{l,r} \ \Rightarrow \ r-l > 0,$
    \item $P_{l,m,r,l'_{1},r'_{1},l'_{2},r'_{2}} \Rightarrow l \leq m < r,$
    \item $P_{l,m,r,l'_{1},r'_{1},l'_{2},r'_{2}} \Rightarrow  Q_{l,m,l'_{1},r'_{1}} \land Q_{m+1,r,l'_{2},r'_{2}}$,
    \item \begin{enumerate}[leftmargin=0.5cm]
        \item[(a)] $R_{l, m} \land S_{l, m} \Rightarrow (P_{(l,m,r)\uparrow\smet{M}_{l,m}} \Rightarrow P_{l,m,r})$
        \item[(b)]  $R_{m, r} \land S_{m, r} \Rightarrow (P_{(l,m,r)\uparrow\smet{M}_{m,r}} \Rightarrow P_{l,m,r})$,
        \end{enumerate}
        where we omit arguments corresponding to the secondary slice, to favour legibility.
\end{itemize}
Finally, the specifications to the various parts of the program scheme are as follows (again, omitting the obvious arguments):
\begin{itemize}
    \item $\prepost{Q \land \ B}{E}{R \wedge S},$
    \item $\prepost{Q \land \ B'}{E'}{R \wedge S},$
    \item $\prepost{Q \land \neg B \land \neg B'}{D}{P},$
    \item $\{ P \land R_{l,m,l'_{1},r'_{1}} \land S_{l,m,l'_{1},r'_{1}} \land \ R_{m+1,r,l'_{2},r'_{2}} \land S_{m+1,r,l'_{2},r'_{2}} \}$\\
    \textbf{C}\\
    $\{ R_{l,r,l',r'} \ \land \  S_{l,r,l',r'} \}$
\end{itemize}

The prime application of the present schema that we have in mind is, as already said, the merge of two slices into a third array. This we proceed to show in the next section.
It is worthwhile pointing out that the merge algorithm to be presented next is specified differently from the one required by the merge sort algorithm given earlier. The one needed in the early section should be implemented by calling the one coming here next.
A copy method becomes adequate to implement the connection between the two merge methods. Incidentally, this copy arises as another application of the present schema, only without the need for a pivot. We omit the details.\\

\section{Merge algorithm, recursively.}
\label{sec:mergeRec}

The instantiation of the preceding schema to originate the merge algorithm is better explained by examining directly the Dafny code (here \emph{without} the necessary proof information):

\hfill

\begin{lstlisting}[basewidth={.55em}]
method {:axiom} merge (a:array<int>, l:int, r:int, 
                       b:array<int>, l':nat, r':nat) 
                       returns (c : array<int>)
 requires 0 <= l <= r <= a.Length
 requires 0 <= l' <= r' <= b.Length
 requires sorted (a[l..r])
 requires sorted (b[l'..r'])
 ensures c.Length == r-l + r'-l'
 ensures sorted (c[..])
 ensures perm (a[l..r]+b[l'.. r'], c[..])
 ensures a[l..r] == old(a[l..r])
 ensures b[l'..r'] == old(b[l'..r'])
 decreases r-l
{
 if r == l { c := new int[r'-l'] ((i) requires 0 <= i < r'-l'
                                     reads b
                                 => b[i+l']) ;
 }
 else{ if r' == l' { c := new int[r-l] ((i) requires 0 <= i < r-l
                                           reads a 
                                        => a[i+l]) ;
 }
 else { 
    var m := (l+r)/2 ;
    var m' := fp (a[m], b, l', r') ;
    
    var d : array<int> := new int[m-l + m'-l'] ;
    d := merge (a, l, m, b, l', m') ;
    var d' : array<int> := new int[r-(m+1) + r'-m'] ;
    d' := merge (a, m+1, r, b, m', r') ;
    
    c := new int[r-l + r'-l']
         ((i) requires 0 <= i < r-l + r'-l'
              reads d 
              reads d' 
              reads a
         => if i < m-l + m'-l'      then d[i]
            else if i > m-l + m'-l' then d'[i - (m-l + m'-l') - 1]
            else a[m]
         ) ;
 }}
}
\end{lstlisting}

\hfill

\begin{figure}[t]
    \centering
    \includegraphics[width=0.85\linewidth]{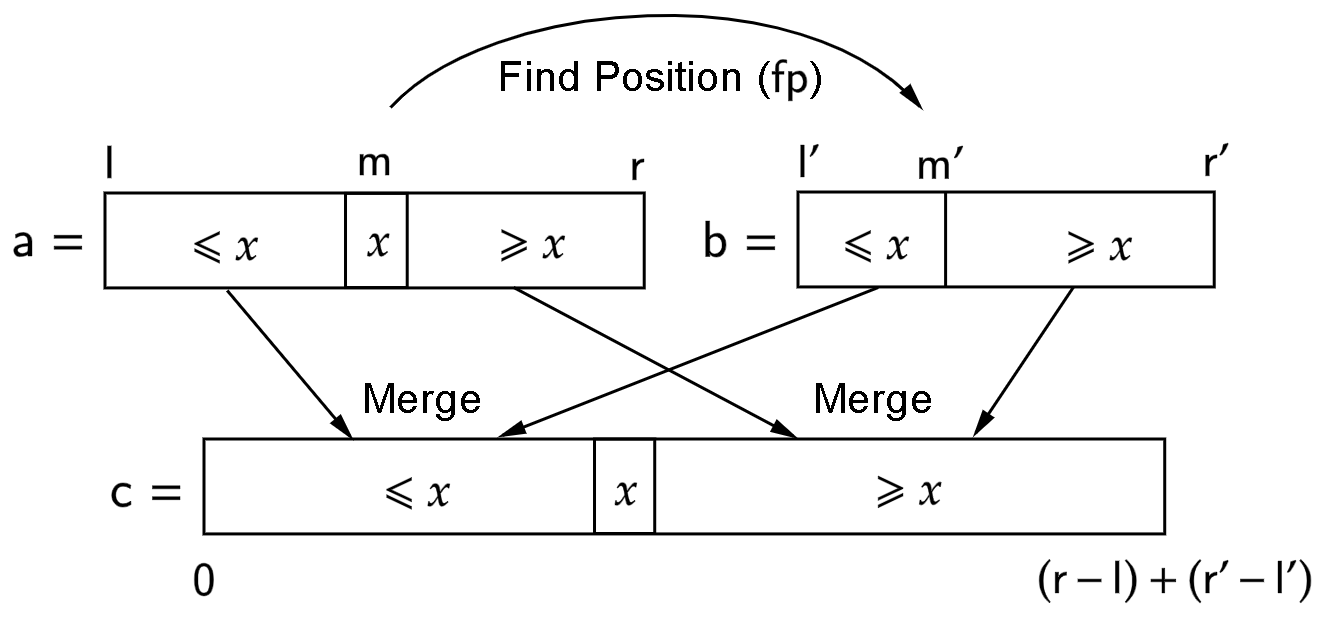}
    \caption{A pictorial depiction of the merge algorithm by the divide-and-conquer strategy.}
    \label{fig:msort-dc}
\end{figure}

Figure~\ref{fig:msort-dc} shows a pictorial depiction of the algorithm idea.  
To begin with, the implementation returns a result array \texttt{c}, and lines 7, 8 and 9 specify this output as expected.
The preconditions in lines 3 and 4 enforce the provisos mentioned in the preceding section as can
be  immediately checked. Lines 10 and 11 ensure the necessary non-interference conditions.

In the body of the method we find, to begin with, the base conditions, called \pred{B} and \pred{B'} in the schema, that here amount just to the participating slices being empty. In these cases we use as methods \met{E} and \met{E'} the creation of \texttt{c} with adequate length and contents, as can be easily read off.

The division phase proceeds by selecting as pivot the value at the middle point of the principal slice (line 23). Then this value is used to partition the secondary slice ---which is accomplished by the function \texttt{fp}. This one ensures that the left hand side of the resulting partition is bounded above by the pivot value, and inversely for the right hand side as can be read from the code below.

Finally, lines 26 to 29 proceed with the recursive calls, just as in the schema and the construction in line 31 is the combine phase.\\

The \texttt{fp} function is the one here below. This code checks as it is.

\hfill

\begin{lstlisting}[basewidth={.55em}]
function fp (x : int, b : array<int>, l : int, r : int) : (m : int)
  reads b
  requires 0 <= l <= r <= b.Length
  requires sorted (b[l..r])
  ensures l <= m <= r
  ensures forall i :: l <= i < m ==> b[i] <= x
  ensures forall i :: m <= i < r ==> x <= b[i]
  decreases r-l
{
  if l == r then l
  else var p := (l+r)/2 ;
           assert l <= p < r ;
       if x == b[p] then p
       else     assert sorted (b[l..p]) && sorted (b[p..r]) 
                by { sub_sorted (b[l..r], 0, r-l, p-l) ;
                     assert (b[l..r])[0..p-l] == b[l..p] ;
                }
           if x < b[p] 
           then fp (x, b, l, p)
           else fp (x, b, p+1, r)
}
\end{lstlisting}

\hfill

Although possible, a solution by divide-and-conquer in this case would not yield a natural procedure. We use the usual tactic of binary search instead. 
Notice the use of lemma \texttt{sub\_sorted} that expresses that sortedness is preserved when taking subsequences.
It remains to exhibit and comment on the verification of \texttt{merge}:

\hfill

\begin{lstlisting}[basewidth={.55em}]
method merge(a:array<int>, l:int, r:int, b:array<int>, 
             l':nat, r':nat) 
 returns (c : array<int>)
 requires 0 <= l <= r <= a.Length
 requires 0 <= l' <= r' <= b.Length
 requires sorted (a[l..r])
 requires sorted (b[l'..r'])
 ensures c.Length == r-l + r'-l'
 ensures sorted (c[..])
 ensures perm (a[l..r]+b[l'.. r'], c[..])
 ensures a[l..r] == old(a[l..r])
 ensures b[l'..r'] == old(b[l'..r'])
 decreases r-l
{
 if r == l { c := new int[r'-l'] ((i) requires 0  <= i < r'-l'
                                      reads b
                                 => b[i+l']) ;
      assert c[..] == b[l'..r'] ;
 }
 else { if r' == l' { c := new int[r-l] ((i) requires 0  <= i < r-l
                                             reads a 
                                      => a[i+l]) ;
      assert c[..] == a[l..r] ;
 }
 else { 
  var m := (l+r)/2 ;
      assert forall i :: l <= i < m ==> a[i] <= a[m] ;
      assert forall i :: m < i < r ==> a[m] <= a[i] ;
  var m' := fp (a[m], b, l', r') ;
      assert forall i :: l' <= i < m' ==> b[i] <= a[m] ;
      assert forall i :: m' <= i < r' ==> a[m] <= b[i] ;
              
      label bfr :
      assert sorted (a[l..m]) &&  sorted (b[l'..m'])
      by {sub_sorted (a[..], l, r, m) ; 
          sub_sorted (b[..], l', r', m') ;
      } 
  var d : array<int> := new int[m-l + m'-l'] ;
  d := merge (a, l, m, b, l', m') ;
  var d' : array<int> := new int[r-(m+1) + r'-m'] ;
  d' := merge (a, m+1, r, b, m', r') ;
     
     calc { 
      a[l..r] == old@bfr(a[l..r]) && b[l'..r'] == old@bfr(b[l'..r']) ;
      ==>
      forall i :: 0 <= i < d.Length ==> (a[l..m]+b[l'..m'])[i] <= a[m] ;
      ==> { perm_leqs (a[l..m]+b[l'..m'], [a[m]], d[..], [a[m]]); }
      forall i :: 0 <= i < d.Length ==> d[i] <= a[m] ;
     }        
     calc {
      a[l..r] == old@bfr(a[l..r]) && b[l'..r'] == old@bfr(b[l'..r']) ;
      ==>
      forall i :: 0 <= i < d'.Length ==> a[m] <= (a[m+1..r]+b[m'..r'])[i] ;
      ==> { perm_leqs ([a[m]], a[m+1..r]+b[m'..r'], [a[m]], d'[..]); }
      forall i :: 0 <= i < d'.Length ==> a[m] <= d'[i] ;
     }
 
  c := new int[r-l + r'-l']
     ((i) requires 0  <= i < r-l + r'-l'
          reads d 
          reads d' 
          reads a
     => if i < m-l + m'-l'      then d[i]
        else if i > m-l + m'-l' then d'[i - (m-l + m'-l') - 1]
        else a[m]
     ) ;
     assert perm (a[l..r]+b[l'.. r'], c[..])
     by { assert c[..] == d[..] + [a[m]] + d'[..] ;
          assert a[l..r] == a[l..m] + [a[m]] + a[m+1..r] ;
          assert b[l'..r'] == b[l'..m'] + b[m'..r'] ;
     }
  }}
} 
\end{lstlisting}

The structure of the executable code is a conditional, with three branches. We therefore have to prove the post-conditions at each branch.

For the first two branches Dafny demands help with the post-condition concerning the permutation relation (line 9); it is then sufficient to assert the value of the result array \texttt{c} (lines 17 and 22).

If we go now directly to the bottom of the code, line 67 expresses the same post-condition at the principal (third) branch of the conditional. The proof in this case consists essentially in expressing again \texttt{c}, this time in terms  
of the results \texttt{d} and \texttt{d'} of the recursive calls, together with the single pivot value \texttt{a[m]} (line 67). The extra assertions in lines 68 and 69 allow to make use of the induction hypotheses.

Moving upwards, we see the block in lines 57 to 65, that creates \texttt{c} and, further up, the recursive calls between lines 37 and 40. What we have in the middle is a proof that paves the way to ascertaining that \texttt{c} is sorted ---which is, by the way, unnecessary to mention later at the very bottom.
The two \texttt{calc} statements are very similar, making use only of a lemma (\texttt{perm\_leqs}) that establishes that ordering between slices (i.e. that a slice is uniformly bounded by another) is preserved by the permutation relation.

It only remains the block from line 25 to 36, i.e. where the partition is performed. What we need to do is prepare the way to establishing the pre-conditions for the recursive calls, which is achieved by the various \texttt{assert} statements therein. The last one (line 33) is obtained by making use of the lemma \texttt{sub\_sorted}, already mentioned.

\section{Divide-and-conquer: from recursion to iteration.}
\label{sec:rec2it}
We are interested now in providing iterative versions of the preceding algorithms, notably of merge sort.
This can be achieved by standard transformations that employ stacks for implementing recursion. However, a more direct approach is also practicable, and we shall presently explore it.

\begin{figure}[H]
    \centering
    \includegraphics[width=0.6\linewidth]{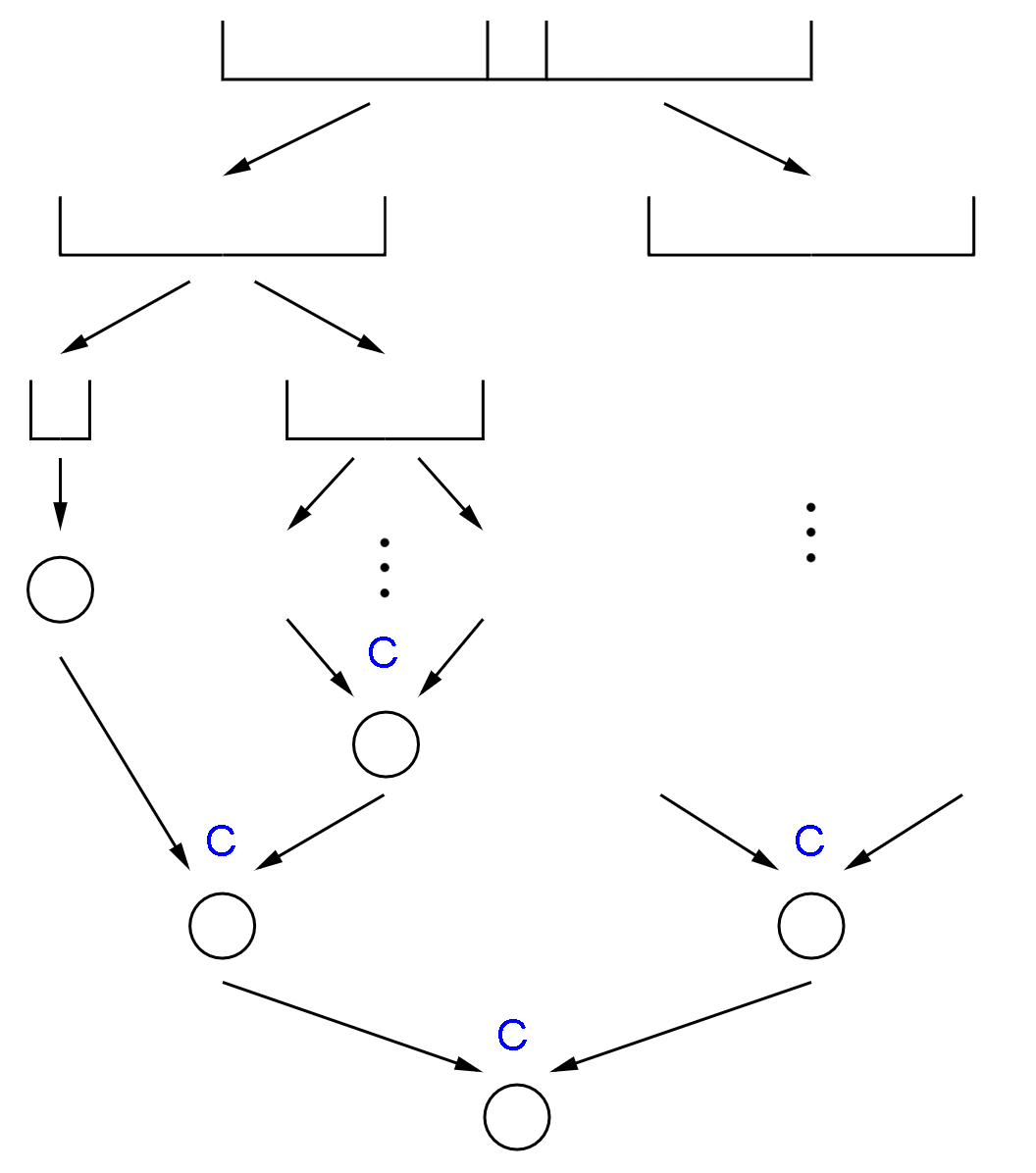}
    \caption{Upper tree of calls and lower tree of combinations in binary divide-and-conquer.}
    \label{fig:rec-tree}
\end{figure}

If we consider the general idea of divide-and-conquer on one array, we are led to the following picture of two trees sharing their border (Figure~\ref{fig:rec-tree}):\\

The upper tree corresponds to the successive phases of division, and the lower tree to those of combination. The circles represent results of the computations, starting right at the border with the base cases.
Assuming sequential computation, one readily notices that a stack is generally necessary to store results of subtrees of the lower tree, in order to later being able to apply the combination phase corresponding to their parent node. 

However, this is not so if the result of the computation consists in updating some global structure (say one residing in the heap).
Indeed, in such cases one can think of the procedure depicted by the two trees as a combination of two iterations, each of them consisting in one pass on the relevant structure for each level of the corresponding tree. The upper tree iteration would therefore traverse the given array once per level, applying the partition procedure repeatedly, whereas the lower tree iteration would do the same for the global structure to be updated, repeating the combination phase on sub-structures.
The upper iteration would proceed until arriving to base slices, whereas the lower one would start from base sub-structures up to the whole.
The idea works at least in the cases in which both partition and combination of sub-structures are independent, as is the case with our non-interference hypotheses.

In the case of merge sort 
there is only one global structure, which is the original array.
Besides, we have a very trivial partition procedure, i.e. just splitting the slices into halves, not affecting the array. Therefore, we can just program the lower iteration, starting from base cases, namely in general, slices of length one, and performing passes on the array that perform merges of consecutive slices. 
This procedure is known as \emph{bottom-up} merge sort and is the one we verify in the next section.

Notice that with respect to quicksort the somehow converse situation obtains. The partition procedure starts in the first pass on the whole array and then repeats on the originated parts until parts of length one are reached all over, in which case the array is sorted, without need for the lower iteration.

\section{Merge sort, iterative version.}
\label{sec:mergesortIt}

We just design the loop depicted in the preceding section. Each pass of it corresponds to a level of the (lower) tree, in which the array is divided into sorted slices of a fixed length, say $s$. 

The initialization yields $s = 1$ (assuming, without much loss, that the given array is non-empty). The end of the iteration happens when $s$ becomes large enough to cover the whole array. And the step consists in merging pairs of consecutive sorted slices, so that $s$ may then double itself. In Dafny:

\begin{lstlisting}[basewidth={.55em}]
method mergeSort (a : array<int>)
 modifies a
 requires a.Length >= 1
 ensures sorted (a[..])
 ensures perm (a[..], old(a[..]))
{
 var s : int ;
 s := 1 ;
 while s < a.Length
  invariant 1 <= s
  invariant forall l :: (0 <= l < a.Length && l%s == 0) 
                         ==> sorted (a[l..min(l+s, a.Length)])
  invariant perm (a[..],old(a[..]))
  decreases a.Length - s
 {
  merges (a, s) ;
  s := 2*s ;
 }
 assert 0%s == 0 ;
}
\end{lstlisting}

Notice the invariant at line 10. It says that the slices starting at indices that are multiples of $s$ are sorted; we take into account that the last slice thereof may be shorter than $s$.

Notice too the rather annoying \texttt{assert} statement at the bottom, which turned out to be necessary for Dafny to establish the post-condition. 

As to \texttt{merges}, it is as follows (without the verification matter):\\

\begin{lstlisting}[basewidth={.55em}]
method {:axiom} merges (a:array<int>, s:int)
 modifies a
 requires 1 <= a.Length
 requires 1 <= s
 requires forall l :: (0 <= l < a.Length && l%s == 0) 
                      ==> sorted(a[l..min(l+s,a.Length)])
 ensures forall l :: (0 <= l < a.Length && l%(2*s) == 0) 
                     ==> sorted(a[l..min(l+2*s,a.Length)])
 ensures perm (a[..],old(a[..]))
{
 var j : int ;
 j := 0 ;
 while j != a.Length
  invariant 1 <= s
  invariant 0 <= j <= a.Length
  invariant j != a.Length ==> j%(2*s) == 0
  invariant forall l :: (0 <= l < j && l%(2*s) == 0) 
                        ==> sorted (a[l..min(l+2*s,a.Length)])
  invariant forall l :: (j <= l < a.Length && l%s == 0) 
                        ==> sorted (a[l..min(l+s,a.Length)])
  invariant perm (a[..], old(a[..]))
  decreases a.Length - j
 {
  mergePair (a, j, s) ;
  j := min (j + 2*s, a.Length) ;
 }
}
\end{lstlisting}

\hfill

It consists simply in one pass on the slices of the array. The main invariant (line 16) is derived from the post-condition at line 6 by the simple expedient of replacing the length of the array by the index \texttt{j} used for iterating.
We need however, the invariant at line 17, which says that the pre-condition at line 4 still holds for the part of the array yet not traversed. 
We are left with merging of two consecutive ordered slices of \texttt{a}:\\

\begin{lstlisting}[basewidth={.55em}]
method {:axiom} mergePair (a : array<int>, l : int, s : int)
 modifies a
 requires 0 <= l < a.Length
 requires s >= 1
 requires sorted (a[l .. min (l+s, a.Length)])
 requires sorted (a[min (l+s, a.Length) .. min (l+2*s, a.Length)])
 ensures sorted (a[l .. min (l+2*s, a.Length)])
 ensures perm (a[..], old(a[..]))
 ensures a[0..l] == old(a[0..l])
 ensures a[min (l+2*s, a.Length)..a.Length] 
         == old(a[min (l+2*s, a.Length)..a.Length])
{
 if (l+s < a.Length){
  var aa := new int[s];
  var s' := min(s,a.Length-(l+s));
  var aa' := new int[s'];
  var a' := new int[s+s'];
  copy(a,l,aa,0,s);
  copy(a,l+s,aa',0,s');
  merge(aa,aa',a');
  copy(a',0,a,l,s+s');
 }
}
\end{lstlisting}

\hfill 

This one uses procedures \texttt{merge} and \texttt{copy}, which are basically routine. Their specifications are shown here below, but for further details please look at the appendix.
\\

\begin{lstlisting}[basewidth={.55em}]
method {:axiom} merge(a:array<int>, b:array<int>, c:array<int>)
  requires sorted(a[..]) && sorted(b[..])
  requires c.Length == a.Length + b.Length 
  modifies c
  ensures sorted(c[..])
  ensures multiset(c[..c.Length]) 
          == multiset(a[..a.Length] + b[..b.Length])
  ensures a[..] == old(a[..]) &&  b[..] == old(b[..])
\end{lstlisting}

\begin{lstlisting}[basewidth={.55em}]
method {:axiom} copy(src:array<int>, i:int, 
                     dest:array<int>, j:int, len:int)
 requires 0 <= i < src.Length && 0 <= j < dest.Length && 0 <= len
 requires src.Length >= i+len && dest.Length >= j+len
 requires src != dest
 modifies dest
 ensures src[..] == old(src[..])
 ensures dest[..j] == old(dest[..j])
 ensures dest[j+len..] == old(dest[j+len..])
 ensures dest[j..j+len] == src[i..i+len]
\end{lstlisting}


\section{Conclusions.}
\label{sec:conclusions}

We have presented full certifications in Dafny of two variants of merge sort ---respectively recursive and iterative---, derived generally from analyses of the divide-and-conquer method that conducted the schemas of programs and proofs. The recursive version includes a (recursive) merge algorithm also obtained by a divide-and-conquer schema, and the iterative one follows from an observation on the tree of calls of the recursive solution and does not employ stacks.

We have tried to explain the most relevant matter concerning the proof code provided to Dafny in order to achieve the full verification. Whereas some of it has come up as surprising because of their obvious nature, the whole bulk has resulted very much manageable. 
It is worthwhile to comment that, at the time of its completion, the present is about the biggest project the authors have attempted using this tool. 
We believe it is a good sign that we have completed it successfully in a rather short period of time. It also needs to be said that the process has been conducted by approaching Dafny only after careful reflection on the matter at hand.

This work is very much of the kind of 
Certezeanu et al's~\cite{QuicksortRevisited}, who present a verification of recursive and iterative versions of quicksort in Dafny, including an original iterative variant based on a stack of pivot positions. While their recursive version is fully verified using Hoare logic and auxiliary lemmas, the verification of the iterative variant remains incomplete. Their work is primarily concerned with reasoning about given algorithm implementations. In contrast, our approach emphasizes the derivation of correct programs from general-purpose divide-and-conquer schemas, with verification naturally integrated into the development process. 

To the best of our knowledge, the present is the first full certification of array sorting by merging in both the recursive and iterative versions, besides they being derived from abstract schemas for the divide-and-conquer method.
In~\cite{Leino:NaturalMS}, Leino and Lucio provide a fully verified implementation of natural mergesort in Dafny, with a strong emphasis on proving both correctness and stability. Their approach uses functional programming over algebraic lists and employs Dafny’s assertional style, including intermediate assertions, ghost code, and structured lemmas. The verification targets a specific functional implementation and highlights Dafny’s expressiveness for reasoning about higher-level properties like stability. In contrast, our work focuses on the derivation and certification of sorting algorithms over arrays, using parameterized divide-and-conquer schemas. 
We treat both recursive and iterative variants within a unified approach, emphasizing modular specifications and non-interference. While we do not address stability explicitly, our goal is to support replicable and transparent correctness derivation, especially for array-based implementations.

Shi et al.~\cite{Shi2008:MPAR} propose a formal derivation method for generating both algorithms and their loop invariants from high-level specifications. Their approach, grounded in the PAR framework and the Radl specification language, is particularly effective for sorting problems like mergesort, which they derive mechanically along with corresponding invariants. The focus is on reducing the mathematical burden of formal development through automated tools and transformation strategies. While we share their goal of deriving verified sorting algorithms, our methodology differs in structure and emphasis: we work directly in Dafny, targeting array-based formulations, and derive both recursive and iterative implementations from a unified divide-and-conquer schema. Our loop invariants arise naturally from the structure of the recursive decomposition, enabling a seamless transition from specification to certified implementation.

Ji et al.~\cite{Ji2024} present AutoLifter, a system for synthesizing divide-and-conquer–like algorithms by decomposing relational specifications into inductively solvable subtasks using component and variable elimination. Their approach focuses on automating synthesis via counterexample-guided techniques, without formal verification. While we share a structural focus on divide-and-conquer, our work differs in intent and methodology: we derive sorting algorithms through schema-guided development with correctness built in from the outset, and provide full formal verification in Dafny, including iterative variants with derived loop invariants.

We would now like to make use of the present formulations for achieving a full certification of parallelized versions
of the algorithms.
Equally, it would be interesting to investigate proofs of complexity bounds (see e.g.~\cite{Morshtein2021}), especially for the parallel versions (see e.g.~\cite{Leino2017}).

We also are interested in completing a similar analysis for quicksort.

\bibliographystyle{ACM-Reference-Format}
\bibliography{refs}

\appendix
\section{Dafny Code}
\label{appendix}

In this appendix we include some of the code referred to in the main article, exhibiting full verification matter. The whole development can be found at: {\em 
\url{https://github.com/josedusol/merge-sort-dafny-verification}}.\\

\begin{lstlisting}[caption={Iterative merging of all consecutive slices.},basewidth={.55em}]
method merges (a : array<int>, s : int)
 modifies a
 requires 1 <= a.Length
 requires 1 <= s
 requires forall l :: (0 <= l < a.Length && l%s == 0) 
                      ==> sorted(a[l..min(l+s,a.Length)])
 ensures forall l :: (0 <= l < a.Length && l%(2*s) == 0) 
                     ==> sorted(a[l..min(l+2*s,a.Length)])
 ensures perm (a[..], old(a[..]))
{
 var j : int ;
 j := 0 ;
 while j != a.Length
  invariant 1 <= s
  invariant 0 <= j <= a.Length
  invariant j != a.Length ==> j%(2*s) == 0
  invariant forall l :: (0 <= l < j && l%(2*s) == 0) 
                         ==> sorted (a[l..min(l+2*s, a.Length)])
  invariant forall l :: (j <= l < a.Length && l%s == 0) 
                         ==> sorted (a[l..min(l+s, a.Length)])
  invariant perm (a[..], old(a[..]))
  decreases a.Length - j
 {          
    assert j%s == 0 by { modn_i (j, s) ;}
    assert sorted (a[min(j+s, a.Length)..min(j+2*s,a.Length)]) 
    by { assert 0 <= j+s < a.Length ==> (j+s)%s == 0 
         by { modn (j, s) ;}
    }
    label mp :
   mergePair (a, j, s) ;
    forall l : int | 0 <= l < min(j+2*s, a.Length) && l%(2*s) == 0
     ensures sorted (a[l..min(l+2*s, a.Length)])
    { 
     if l < j { 
      assert 0 <= l < l+2*s <= j < a.Length 
      by {mod2n_ii (j, l, 2*s) ;}
      assert a[l..min(l+2*s, a.Length)]
             ==(a[0..j])[l..min(l+2*s, a.Length)] 
      by { sub_subseq (a[..], 0, j, l, min(l+2*s, a.Length)) ;}
      assert old@mp((a[0..j])[l..min(l+2*s,a.Length)]) 
             == old@mp(a[l..min(l+2*s,a.Length)])
      by { sub_subseq(old@mp(a[..]), 0, j, l, min(l+2*s, a.Length));}
     } else {mod2n_iii (j, l, 2*s) ;}
    }
    calc {
        j + 2*s < a.Length ;
    ==>
        j < a.Length ;
    ==>  {assert j < a.Length ==> j != a.Length ==> j%(2*s) == 0 ;}
        j%(2*s) == 0 ;
    ==>         {mod2n (j, 2*s) ;}
        (j + 2*s)%(2*s) == 0 ;
    } 
   j := min (j + 2*s, a.Length) ;
   forall l : int | j <= l < a.Length && l%s == 0
    ensures sorted (a[l .. min(l+s, a.Length)])
   { assert a[l..min(l+s, a.Length)] 
            == (a[j..a.Length])[l-j..min(l+s, a.Length)-j] 
     by { sub_subseq (a[..], j, a.Length, l, min(l+s, a.Length)) ;
   }
     assert old@mp((a[j..a.Length])[l-j..min(l+s,a.Length)-j]) 
            ==old@mp(a[l..min(l+s,a.Length)]) 
     by{ sub_subseq(old@mp(a[..]),j,a.Length,l,min(l+s,a.Length)); }
   }
 }
}
\end{lstlisting}

\hfill\\\\\\\\\\

\begin{lstlisting}[caption={Merge of two consecutive slices.},basewidth={.55em}]
method mergePair (a : array<int>, l : int, s : int)
 modifies a
 requires 0 <= l < a.Length
 requires s >= 1
 requires sorted (a[l .. min (l+s, a.Length)])
 requires sorted (a[min (l+s, a.Length) .. min (l+2*s, a.Length)])
 ensures sorted (a[l .. min (l+2*s, a.Length)])
 ensures perm (a[..], old(a[..]))
 ensures a[0..l] == old(a[0..l])
 ensures a[min (l+2*s,a.Length)..a.Length] 
         == old(a[min (l+2*s, a.Length)..a.Length])
{
 if (l+s < a.Length){
  var aa := new int[s];
  var s' := min(s,a.Length-(l+s));
  var aa' := new int[s'];
  var a' := new int[s+s'];
          
  copy(a,l,aa,0,s);
  copy(a,l+s,aa',0,s');
   assert old(a[l..l+s+s'])==aa[0..s]+aa'[0..s'];
  merge(aa,aa',a');
  copy(a',0,a,l,s+s');
   assert a[..] == a[0..l]+a[l..l+s+s']+a[l+s+s'..a.Length];
   assert old(a[..]) == old(a[0..l]+a[l..l+s+s']+a[l+s+s'..a.Length]);
 }
}
\end{lstlisting}

\hfill

\begin{lstlisting}[caption={Merge of two arrays.},basewidth={.55em}]
method merge(a:array<int>, b:array<int>, c:array<int>)
 requires sorted(a[..]) && sorted(b[..])
 requires c.Length == a.Length + b.Length 
 modifies c
 ensures sorted(c[..])
 ensures multiset(c[..c.Length]) 
         == multiset(a[..a.Length] + b[..b.Length])
 ensures a[..] == old(a[..]) &&  b[..] == old(b[..])
{
 var i, j, k := 0, 0, 0;
 while i < a.Length && j < b.Length
  invariant 0 <= i <= a.Length && 0 <= j <= b.Length && k == i + j 
  invariant sorted(c[..k])
  invariant forall m, n :: 0 <= m < k && i <= n < a.Length ==> c[m] <= a[n]
  invariant forall m, n :: 0 <= m < k && j <= n < b.Length ==> c[m] <= b[n]
  invariant a[..] == old(a[..]) && b[..] == old(b[..])
  invariant multiset(c[..k]) == multiset(a[..i] + b[..j])
 {
  if a[i] <= b[j] {
   c[k] := a[i];
   i := i + 1;
  } else {
   c[k] := b[j];
   j := j + 1;
  }
  k := k + 1;
 }
   
 if (i < a.Length) { 
  while i < a.Length
   invariant 0 <= i <= a.Length  &&  0 <= k <= c.Length &&  k == i + j
   invariant sorted(c[..k])
   invariant forall m, n :: 0 <= m < k && i <= n < a.Length ==> c[m] <= a[n] 
   invariant a[..] == old(a[..]) && b[..] == old(b[..]) 
   invariant multiset(c[..k]) == multiset(a[..i]) + multiset(b[..j])  
  {
   c[k] := a[i];
   i, k := i+1, k+1;
  }
 }
 if (j < b.Length) { 
  while j < b.Length
   invariant 0 <= j <= b.Length  &&  0 <= k <= c.Length  &&  k == i + j
   invariant sorted(c[..k])
   invariant forall m, n :: 0 <= m < k && j <= n < b.Length ==> c[m] <= b[n]
   invariant a[..] == old(a[..]) && b[..] == old(b[..]) 
   invariant multiset(c[..k]) == multiset(a[..i]) + multiset(b[..j])
  {
   c[k] := b[j];
   j, k := j+1, k+1;
  }
 }
}
\end{lstlisting}
\end{document}